\documentclass{IEEEtran}
\usepackage{siunitx}
\usepackage{cite}
\usepackage{amsmath,amssymb,amsfonts}
\usepackage{algorithmic}
\usepackage{graphicx}
\usepackage{textcomp}
\usepackage{color,soul}
\usepackage{dblfloatfix}    % To enable figures at the bottom of page
\usepackage{tabularx,booktabs}
\def\BibTeX{{\rm B\kern-.05em{\sc i\kern-.025em b}\kern-.08em
    T\kern-.1667em\lower.7ex\hbox{E}\kern-.125emX}}

\usepackage{dcolumn,booktabs}
\newcolumntype{d}[1]{D{.}{.}{#1}}
\begin{document}
\title{On the Application of Numerical Methods to Hall\'en's Equation: The Case of a Lossy  Medium}

\author{Christos Mystilidis\thanks{C. Mystilidis and G. Fikioris are with the School of Electrical and Computer Engineering, National Technical University of Athens, GR 157-73 Zografou, Athens, Greece (e-mail: christosmy96@hotmail.com, gfiki@ece.ntua.gr).}, Panagiotis J. Papakanellos\thanks{P. J. Papakanellos is with the Department of Aeronautical Sciences, Hellenic Air Force Academy, Dekelia Air Force Base, Athens 13671, Greece (email: papakanellosp@yahoo.gr).},\\ George Fikioris, \IEEEmembership{Senior Member, IEEE}, and Themistoklis K. Mavrogordatos \thanks{Th. K. Mavrogordatos is with the Department of Physics, Stockholm University, SE-106 91, Stockholm, Sweden (email: themis.mavrogordatos@fysik.su.se)}}
%\thanks{This paragraph of the first %footnote will contain the date on 
%which you submitted your paper for review. %It will also contain support 
%information, including sponsor and %financial support acknowledgment. For 
%example, ``This work was supported in part %by the U.S. Department of 
%Commerce under Grant BS123456.'' }

\maketitle

\begin{abstract}
A previous paper analyzed in detail the difficulties associated with the application of numerical methods to Hall\'en's integral equation with the approximate kernel for the case of a lossless surrounding medium. The present paper extends to the case where the medium is conducting and points out similarities and differences between the two cases. Our main device is an analytical/asymptotic study of the antenna of infinite length. 
\end{abstract}

\begin{IEEEkeywords}
Antenna theory, antennas in matter, integral equations, moment methods, wire antennas
\end{IEEEkeywords}

\section{Introduction}
\label{sec:introduction}

The purpose of the 2001 paper \cite{fikioris1} is to analyze the difficulties associated with the numerical solution of the usual (Hall\'en and Pocklington)  integral equations for a thin-wire transmitting antenna. In \cite{fikioris1}, the antenna is isolated in a lossless medium, is perfectly conducting, and is center-driven by a delta-function generator. For the case of the well-known approximate (also called reduced) kernel, the main difficulties arise from the fact that neither of the said integral equations has a solution; the important issue of ``nonsolvability'' is discussed in detail in  \cite{fikioris1, wu, fikioris2}. Since 2001, the  analysis of \cite{fikioris1} has been extended in a number of directions, including
the so-called ``extended thin-wire kernel'' \cite{papakanellos1}, feeds other than the delta-function generator \cite{fikioris3,fikioris4,tastsoglou1, tastsoglou2}, loop antennas \cite{fikioris5}, \cite{mckinley}, a similar equation of electrostatics \cite{fikioris6}, as well as antennas with finite conductivity, including carbon-nanotube antennas \cite{fikioris7}. Furthermore, the analysis of \cite{fikioris1} forms the foundation for the development of an   easy-to-apply technique  \cite{fikioris6,papakanellos2,fikioris8,fikioris9,mavrogordatos} that is an \textit{a posteriori} remedy for the most important difficulties.

The present article extends results of \cite{fikioris1} toward a different direction, specifically to the case where the medium surrounding the thin-wire antenna is conducting. A standard general reference for such antennas is \cite{kingsmith}, while \cite{richmond} contains moment-method analyses, and the recent paper \cite{hanson3} is a pertinent application to carbon nanotubes.  Section II contains
straightforward numerical results that explicate the difficulties associated with moment-method solutions. Then, in Section III, we explain the findings of Section II by means of an analytical study of the much simpler antenna of infinite length. Most of the derivations pertaining to Section III are in the Appendix, whose contents parallel material in \cite{fikioris1}. 

The approximate kernel, which is a much simpler version of the so-called exact kernel, is extensively used alongside its ``exact'' counterpart, featuring in the majority of modern antenna textbooks; see, e.g., \cite{new1, new2, new3, new4}. Furthermore, extensions of the integral equations with the approximate kernel apply to more involved ``real-life'' antenna configurations, which are typically dealt with by standard antenna-analysis software such as the popular Numerical Electromagnetics Code (NEC) \cite{new5}. Let us add that the approximate kernel enjoys wide popularity in the literature dealing with Hall\'{e}n's and Pocklington's equations for carbon nanotube antennas (CNTs). Characteristic examples include  \cite{new6,new7,new8},  where the approximate kernel is used exclusively, while in other works both kernels are employed (e.g., \cite{new9,new10}).

It is perhaps to be expected that the difficulties we find in this paper are quite different from the difficulties (discussed recently in \cite{fikioris7}, \cite{mavrogordatos}) that arise in the case where the antenna itself is an imperfect conductor \cite{hanson1}, \cite{hanson2}. On the other hand, it is probably surprising that our central analytical result for the infinite antenna---eqn. (\ref{eq:asymptotic}) below---appears very similar to the corresponding result of \cite{fikioris1} for the case of a lossless surrounding medium; compare our (\ref{eq:asymptotic}) to eqn. (38) of \cite{fikioris1}. The predictions of the two results, however, are not the same. This is why our derivations and discussions focus on the similarities/differences with \cite{fikioris1}.

We close this Introduction by listing the main  conclusions from \cite{fikioris1}: For the case of a lossless surrounding medium and for $N\gg 1$, where $2N+1$ is the number of subdomain basis functions, the moment-method solutions for the current $I(z)$ exhibit large and unphysical oscillations near the driving point at $z=0$. These oscillations specifically occur in the imaginary part $\textrm{Im}\{I(z)/V\}$ ($V$ is the driving voltage at $z=0$), while the real part presents no such oscillations. Additional oscillations, in both the real and imaginary parts, occur near the antenna endpoints.  The oscillations are not due to roundoff or matrix-ill-conditioning effects. Rather, they are due to the aforementioned ``nonsolvability,'' discussed particularly clearly in  \cite{wu}. Condition numbers are large and thus magnify roundoff errors \cite{fikioris3}, but these are separate issues.
The near field associated with the oscillating currents \cite{fikioris6,papakanellos2,fikioris8,fikioris9,mavrogordatos} exhibits a superdirective-type behavior, oscillating up to a radial distance $\rho$ equal to the antenna radius $a$. After $\rho=a$, the oscillations cease and the field is similar to that produced by the current satisfying the exact integral equation. This similarity, in fact, forms the basis for the \textit{a posteriori} remedy mentioned previously. 

\begin{figure}[t!]
    \centering
    \includegraphics[width=\columnwidth]{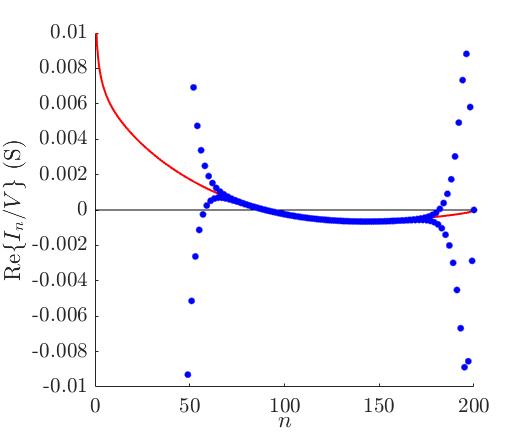}
    \caption{Real part of $I_n/V$ (dots) as calculated by Galerkin's method;  $\epsilon=\epsilon_0$, $\mu=\mu_0$,   {${\omega=2\pi\times 5\times10^8\text{ rad/s}}$}, $\gamma=0.1\text{ S/m}$,  $h=0.25\lambda$, $a=0.007022\lambda$, and $N=200$. Some values near $n=0$ are out of scale and are not shown. The solid lines are corresponding results with the exact kernel.}
    \label{fig:my_label}
\end{figure}
\section{Difficulties associated with finite antenna}

The current $I(z)$ on a wire antenna center-driven by a delta-function generator satisfies Hall\'en's equation \cite{kingsmith,kingharrison,popovic}
\begin{equation}
\label{eq:hallensquation}
\int_{-h}^h K(z-z^{\prime}) I(z^{\prime})\,dz^{\prime}=\frac{iV}{2\zeta_c}\sin k_c|z|+C\cos k_cz,\ |z|<h
\end{equation}
where
\begin{equation}
\label{eq:kernel}
K(z)=\frac{1}{4\pi}\frac{\exp\left(ik_c\sqrt{z^2+a^2}\right)}{\sqrt{z^2+a^2}}
\end{equation}
In (\ref{eq:hallensquation}) and (\ref{eq:kernel}), $K(z)$ is the approximate kernel, $2h$ and $2a$ are the antenna length and diameter, $C$ is a constant to be determined from $I(\pm h)=0$ (see \cite{kingsmith} for discussions on this condition), and an
 $e^{-i\omega t}$ time dependence is assumed. The complex parameters $k_c$ and $\zeta_c$ are 
 \begin{equation}
 \label{eq:parameters}
  k_c=\omega\sqrt{\mu \epsilon_c},\quad
 \zeta_c=\sqrt{\frac{\mu}{\epsilon_c}},\quad
 \epsilon_c=\epsilon+i\frac{\gamma}{\omega}
 \end{equation}
in which the real paramaters $\epsilon$, $\mu$, and $\gamma$ are the permittivity, permeability, and conductivity of the surrounding medium. While standard in the literature, we stress that (\ref{eq:hallensquation}) is approximate due to current leakage into the lossy surrounding medium \cite{kingharrison}.
In (\ref{eq:hallensquation}) and (\ref{eq:kernel}), the differences from \cite{fikioris1} (whose figures pertain to a lossless medium with $\gamma=0$) appear in $\zeta_c$ and $k_c$, which now have nonzero imaginary parts, assumed to satisfy $\textrm{Im}\{k_c\}>0$ and $\textrm{Im}\{\zeta_c\}<0$.

Let $\epsilon=\epsilon_0$, $\mu=\mu_0$, $\lambda=2\pi c/\omega$, $c=3\times 10^8$ m/s,  {${\omega=2\pi f=2\pi\times 5\times10^8\text{ rad/s}}$}, and $\gamma=0.1\text{ S/m}$. At the aforementioned frequency $f$, our chosen value of $\gamma$ corresponds to damp native soil from Kirtland Air Force Base (and is also close to the value for wet Belen soil) \cite{patitz}. Also, let $h=0.25\lambda$, $a=0.007022\lambda$, and $N=200$; these three parameters are the same as in Figs. 1 and 2 of \cite{fikioris1}, with $2N+1$ being the number of pulse basis functions. When applied to (\ref{eq:hallensquation}),   Galerkin's method gives the results shown in Figs. 1 and 2, where  the abscissa $n$ denotes the basis-function number, with $n=0$ at the driving point $z=0$. Corresponding results with the exact kernel are also shown. Our Galerkin's method is identical to the lossless case and is described in detail in \cite{fikioris1}.  The difference with the corresponding figures in \cite{fikioris1} is immediately apparent: Oscillations near the driving point occur not only in  $\textrm{Im}\{I(z)/V\}$, but also in the real part $\textrm{Re}\{I(z)/V\}$ (in fact, we chose  $\gamma$ in order for the real part to exhibit noticeable oscillations). In the next section, we explain this difference, quantitatively, by appealing to the infinite antenna.

\begin{figure}[t]

    \vspace{1.5mm}

    \centering
    \includegraphics[width=1.01\columnwidth]{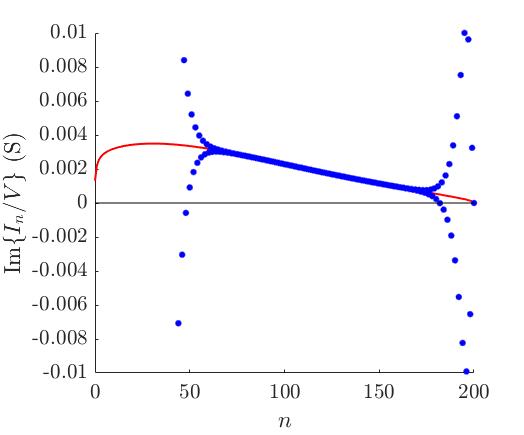}
    \vspace{-1.8em}
    \caption{ Like Fig. 1, but for imaginary part of $I_n/V$.}
    \label{fig:my_label}
\end{figure}
\section{Explanations via infinite antenna}

The infinite-antenna current $I^{(\infty)}(z)$ satisfies the well-known, analogous to (\ref{eq:hallensquation}) integral equation
\begin{equation}
\label{eq:hallensquationinf}
\int_{-\infty}^{\infty} K(z-z^{\prime}) I^{(\infty)}(z^{\prime})\,dz^{\prime}=\frac{V}{2\zeta_c}e^{ik_c|z|},\ -\infty<z<\infty
\end{equation}
(cf. \cite{fikioris1}, \cite{wu}), whose nonsolvability is due to the exponential smallness of the Fourier transform $\bar{K}(\zeta)$ of the approximate kernel $K(z)$, see eqn. (\ref{eq:kernel_asymptotic}) of the Appendix.\footnote{When the exact kernel is used, the corresponding integral equation is solvable and, in fact, one can explicitly calculate the associated fields inside and outside the tube \cite{wu}.} To apply Galerkin's method with pulse functions to (\ref{eq:hallensquationinf}), we set
\begin{equation}
\label{approximatesolution}
    I^{(\infty)}(z)\cong\sum_{n=-\infty}^{\infty}I^{(\infty)}_nu_n(z), \quad -\infty<z<\infty
\end{equation}
where $u_n(z)$ ($n=0,\pm 1,\ldots$) are the pulse basis functions of \textit{finite} width $z_0$, with $u_0(z)$ centered at $z=0$. Substitute (\ref{approximatesolution}) into (\ref{eq:hallensquationinf}), multiply by $u_l(z)$, and integrate with respect to $z$ to obtain the system of equations
\begin{equation}
\label{eq:toeplitz}
    \sum_{n=-\infty}^{\infty}A_{l-n}I^{(\infty)}_n=B_l, \quad l=0,\pm1,\pm2,...
\end{equation}
for the unknown basis-function coefficients $I^{(\infty)}_n$, where $A_l$ and $B_l$ are determined in the Appendix.

The key idea is that the system (\ref{eq:toeplitz}) is doubly infinite and Toeplitz and can therefore be solved exactly for nonzero $z_0$ using Fourier series, i.e., by application of the usual process of discrete deconvolution. This is done in the Appendix. Then, the Appendix determines the asymptotic behavior of $I^{(\infty)}_n$ subject to the conditions
\begin{equation}
\label{eq:conditions}
\frac{z_0}{a}\ll 1,\quad \frac{nz_0}{a}=O(1), \quad |k_c|z_0 \ll 1
\end{equation}
The first two conditions are the same as in \cite{fikioris1}, while the third one is new. The final asymptotic result is  
\begin{equation}
\label{eq:asymptotic}
    \begin{split}
        I^{(\infty)}_{n} \sim     &-i\frac{V}{\zeta_c}\frac{\pi^3}{32\sqrt{2}}k_c\,z_0\sqrt{\frac{z_0}{a}}(-1)^n \\
        &\times\exp{\left(\frac{a\pi}{z_0}\right)}
        \frac{1}{\cosh{\left(\frac{\pi}{2}\frac{z_0}{a}n\right)}} \\
        &\times\left[1-\frac{5}{2\pi}\frac{z_0}{a}+\frac{5}{4}n\left(\frac{z_0}{a}\right)^2\tanh{\left(\frac{\pi}{2}\frac{z_0}{a}n\right)}\right]
    \end{split}
\end{equation}
As already mentioned in our Introduction, (\ref{eq:asymptotic}) is very similar to the corresponding equation in \cite{fikioris1}, namely eqn. (38) of \cite{fikioris1}. In fact, the only difference is that the real parameters $k$ and $\zeta_0$ of \cite{fikioris1} are replaced by the complex parameters $k_c$ and $\zeta_c$. Since (\ref{eq:parameters}) gives
\begin{equation}
\frac{k_c/\zeta_c}{k/\zeta_0}=\frac{\omega\epsilon_c}{\omega\epsilon}=1+i\frac{\gamma}{\omega\epsilon}
\end{equation}
we can divide (\ref{eq:asymptotic}) by eqn. (38) of \cite{fikioris1} to obtain 
\begin{equation}
\label{eq:asymptotic-simple}
\frac{I^{(\infty)}_{n}}{V} \sim (1+i\tan\delta)\, \frac{I^{(\infty)}_{n, \textrm{lossless}}}{V}
\end{equation}
where $\tan\delta=\gamma/(\omega\epsilon)$ is the usual loss tangent of the surrounding medium and $I^{(\infty)}_{n, \textrm{lossless}}$ is the corresponding quantity for the lossless case (i.e., the basis-function coefficient determined asymptotically in \cite{fikioris1}).
While simple, we stress that (\ref{eq:asymptotic-simple})---which is asymptotic---was not expected beforehand.  No relation similar to (\ref{eq:asymptotic-simple}) holds far from the driving point, or when the pulse width $z_0$ is large. In other words, (\ref{eq:asymptotic-simple}) is valid only  subject to (\ref{eq:conditions}), only describes behavior near $z=0$, and is not valid far from $z=0$.

As $I^{(\infty)}_{n, \textrm{lossless}}/V$ is purely imaginary, (\ref{eq:asymptotic}) and (\ref{eq:asymptotic-simple}) predict oscillations in both the real and the imaginary parts, with a ratio asymptotically equal to $-\tan\delta$, and with oscillations in the imaginary part asymptotically the same as in the lossless case. For parameters as in Figs. 1 and 2, Table~I shows that the first few values of $I_n/V$ are close to the values of $I^{(\infty)}_{n}/V$ obtained from (\ref{eq:asymptotic}). Table~I also verifies (compare to Table~II of \cite{fikioris1}) that $\gamma$  has negligible influence on the oscillations in the imaginary part.  Our Table~I  (as well as many other similar results we have obtained) indicates that the oscillating values obtained via Galerkin's method for the finite antenna can be \textit{quantitatively} estimated from our analytical/asymptotic results for the infinite antenna. We stress that our Table~I is representative of what occurs within a wide range of parameter values, as long as the parameters satisfy (\ref{eq:conditions}).

\begin{table*}[ht]
\caption{Comparison of the first 32 values of the complex $I_n/V$ derived via Galerkin's method (for the finite antenna) to corresponding results obtained via (\ref{eq:asymptotic}) (for the infinite antenna). The parameters ($h/\lambda$, $a/\lambda$, etc.) are those of Figs. 1 and 2.}
\label{table 1}
\centering
\begin{tabular}{c S[table-format=1.2e1, retain-zero-exponent=true] S[table-format=1.2e1, retain-zero-exponent=true] S[table-format=1.2e1, retain-zero-exponent=true] S[table-format=1.2e1, retain-zero-exponent=true] c S[table-format=1.2e1, retain-zero-exponent=true] S[table-format=1.2e1, retain-zero-exponent=true] S[table-format=1.2e1, retain-zero-exponent=true] S[table-format=1.2e1, retain-zero-exponent=true]}
\toprule
 $n$ & $\text{Re}\{I_{n}/V\}$ & $\text{Re}\Big\{ I_{n}^{(\infty)}/V\Big\}$ & $\text{Im}\{I_{n}/V\}$ & $\text{Im}\Big\{I_{n}^{(\infty)}/V\Big\}$ & $n$ & $\text{Re}\{I_{n}/V\}$ & $\text{Re}\Big\{ I_{n}^{(\infty)}/V\Big\}$ & $\text{Im}\{I_{n}/V\}$ & $\text{Im}\Big\{I_{n}^{(\infty)}/V\Big\}$  \\
\midrule
0 & 9.56e2 & 8.95e2 & -2.66e2 & -2.49e2 & 16 & 3.54e1 & 3.58e1 & -9.86e0 & -9.95e0    \\
1 & -9.23e2 & -8.72e2 & 2.57e2 & 2.42e2 & 17 & -2.77e1 & -2.77e1 & 7.72e0 & 7.73e0  \\
2 & 8.37e2 & 8.07e2 & -2.33e2 & -2.25e2 & 18 & 2.17e1 & 2.16e1 & -6.04e0 & -6.00e0 \\
3 & -7.23e2 & -7.14e2 & 2.01e2 & 1.99e2 & 19 & -1.70e1 & -1.67e1 & 4.73e0 & 4.65e0\\
4 & 6.02e2 & 6.08e2 & -1.68e2 & -1.69e2 & 20 & 1.33e1 & 1.30e1 & -3.70e0 & -3.61e0  \\
5 & -4.90e2 & -5.03e2 & 1.36e2 & 1.40e2 & 21 & -1.04e1 & -1.00e1 & 2.91e0 & 2.79e0 \\
6 & 3.93e2 & 4.07e2 & -1.09e2 & -1.13e2 & 22 & 8.17e0 & 7.78e0 & -2.27e0 & -2.16e0 \\
7 & -3.13e2 & -3.26e2 & 8.70e1 & 9.06e1 & 23 & -6.39e0 & -6.02e0 & 1.78e0 & 1.67e0  \\
8 & 2.47e2 & 2.58e2 & -6.88e1 & -7.12e1 & 24 & 5.01e0 & 4.66e0 & -1.39e0 & -1.30e0  \\
9 & -1.95e2 & -2.03e2 & 5.41e1 & 5.65e1 & 25 & -3.92e0 & -3.60e0 & 1.10e0 & 1.00e0  \\
10 & 1.53e2 & 1.59e2 & -4.25e1 & -4.44e1 & 26 & 3.08e0 & 2.78e0 & -8.52e-1 & -7.74e-1 \\
11 & -1.20e2 & -1.25e2 & 3.34e1 & 3.47e1 & 27 & -2.40e0 & -2.15e0 & 6.73e-1 & 5.98e-1  \\
12 & 9.40e1 & 9.74e1 & -2.62e1 & -2.71e1 & 28 & 1.89e0 & 1.66e0 & -5.21e-1 & -4.61e-1\\
13 & -7.37e1 & -7.60e1 & 2.05e1 & 2.11e1 & 29 & -1.47e0 & -1.28e0 & 4.14e-1 & 3.56e-1 \\
14 & 5.77e1 & 5.92e1 & -1.61e1 & -1.65e1 & 30 & 1.16e0 &  9.88e-1 & -3.18e-1 & -2.75e-1  \\
15 & -4.52e1 & -4.60e1 & 1.26e1 & 1.28e1 & 31 & -9.01e-1 & -7.62e-1 & 2.55e-1 & 2.12e-1  \\
\bottomrule
\end{tabular}
\end{table*}

\section{Conclusions, extensions, future work}

The main difficulty associated with the moment-method solutions of the usual thin-wire integral equations with the approximate kernel is known from \cite{fikioris1}: For a sufficiently large number of basis functions, the imaginary part $\textrm{Im}\{I(z)/V\}$ is very large and oscillates rapidly. In the present paper, we extended this result to the case where the surrounding medium is imperfectly conducting. It was found that oscillations occur in both the real and the imaginary parts. As in the lossless case, the oscillations are not due to roundoff errors or to matrix- ill-conditioning effects: while important (e.g. the admittance matrix pertaining to Figs. 1 and 2 and Table I has a condition number equal to $5 \times 10^8$), such issues are completely separate. 

As in \cite{fikioris1}, our analytical/asymptotic results for the infinite antenna help us understand the behavior of the numerical solutions for the finite one. Furthermore, many of the extensions and remarks of \cite{fikioris1} continue to hold in the present case. For example, since $(2N+1)z_0=2h$ in the finite antenna, the condition  $z_0\gg a$ for oscillations near the driving point [see (\ref{eq:conditions})] translates to $N\gg h/a$ for the finite antenna. Also, oscillations occur with different basis and testing functions; these, however, can give rise to  different asymptotic formulas. Many of our results carry over, without modification, to Pocklington's equation. Finally, our analytical/asymptotic study is only relevant to oscillations near the driving point; for a study of oscillations near the endpoints (see our two figures), one must solve a Wiener-Hopf sum equation and (as in the lossless case \cite{fikioris1}) this seems difficult to carry out.

We have additionally applied our methods to the so-called ``extended thin-wire kernel'' (see \cite{papakanellos1} for the lossless case). Our main conclusions continue to hold, while the essential benefits of the extended kernel (milder and slower oscillations) were verified via extensive numerical experiments.  

Future work will focus on the  \textit{a posteriori} remedy mentioned in our Introduction. The detailed understanding obtained via (\ref{eq:asymptotic-simple}) is expected (as in the lossless case \cite{fikioris8}, \cite{fikioris9}) to greatly facilitate this   study.

\section*{Acknowledgment}

We thank C. Arvanitis and M. Bagakis for helping with typesetting.

\section{Appendix}

\newcounter{mytempeqncnt}
\begin{figure*}[!b]% ensure that we have normalsize text
\normalsize% Store the current equation number.
\setcounter{mytempeqncnt}{\value{equation}}% Set the equation number to one less than the one% desired for the first equation here.% The value here will have to changed if equations% are added or removed prior to the place these% equations are referenced in the main text.
\setcounter{equation}{17}
\hrulefill% The spacer can be tweaked to stop underfull vboxes.

\begin{equation}
\label{eq:big_1}
I^{(\infty)}_{n} \sim \frac{(-1)^n}{4\pi z_0}\int_{0}^{\pi}\frac{\bar{B}(\pi-\phi)/\cos^2{\frac{\phi}{2}}}
    {\bar{K}\left(\frac{\pi-\phi}{z_0}\right)/(\pi-\phi)^2+\bar{K}\left(\frac{\pi+\phi}{z_0}\right)/(\pi+\phi)^2}\cos{n\phi}d\phi
\end{equation}
%
%\begin{equation}
%\label{eqn_dbl_y}
%\lim_{z_0\to0}\textrm{Re}{\left\{\frac{I^{(\infty)}_{n}}{V}\right\}}=
%    \frac{4k_c}{\pi\zeta_0}\int_{0}^{k_c}\frac{J_0\left(a\sqrt{k_c^2-\zeta^2}\right)\cos{(\zeta nz_0})}
%    {(k_c^2-\zeta^2)\left[J^2_0\left(a\sqrt{k_c^2-\zeta^2}\right)+Y^2_0\left(a\sqrt{k_c^2-\zeta^2}\right)\right]}d\zeta
%\end{equation}
% Restore the current equation number.
\setcounter{equation}{\value{mytempeqncnt}}% IEEE uses as a separator

\vspace*{0pt}
\end{figure*}

As long as $k$ and $\zeta$ are replaced by $k_c$ and $\zeta_c$, eqns. (20) and (8) of \cite{fikioris1} continue to hold, giving a closed-form expression for $B_l$ and a single-integral expression for $A_l$.  To solve the infinite system (\ref{eq:toeplitz}), introduce the Fourier series
\begin{equation}
\label{eq:fourier_series}
    \begin{split}
        &\bar{A}(\theta)=\sum_{l=-\infty}^{\infty}A_le^{il\theta}, \quad \bar{B}(\theta)=\sum_{l=-\infty}^{\infty}B_le^{il\theta} \\
        &\bar{I}(\theta)=\sum_{l=-\infty}^{\infty}I^{(\infty)}_le^{il\theta}
    \end{split}
\end{equation}
and the use the convolution theorem to obtain
\begin{equation}
\label{eq:fourierexp}    
    I^{(\infty)}_n=\frac{1}{2\pi}\int_{-\pi}^{\pi}\frac{\bar{B}(\theta)}{\bar{A}(\theta)}
                                                e^{-in\theta}d\theta =\frac{1}{\pi}\int_{0}^{\pi}\frac{\bar{B}(\theta)}{\bar{A}(\theta)}
                             \cos({}n\theta)d\theta 
\end{equation}
The Fourier series for $\bar{B}(\theta)$ converges by the aforementioned closed-form expression for $B_l$ and the condition $\textrm{Im}\{k_c\}>0$ of Section II. Direct summation\footnote{This is somewhat simpler than the corresponding step in \cite{fikioris1}, where one must additionally assume that $\textrm{Im}\{k\}>0$. As a result, the integrals corresponding to those in (\ref{eq:fourierexp}) have indented integration contours.} then gives
\begin{equation}
\label{eq:btheta}
    \bar{B}(\theta)=-\frac{iV}{\zeta_c}\frac{2}{k_c}\sin^2{\frac{k_cz_0}{4}}\frac{\cos{\frac{k_cz_0}{2}}+\cos^2{\frac{\theta}{2}}}{\sin{\frac{\theta+k_cz_0}{2}}\sin{\frac{\theta-k_cz_0}{2}}}
\end{equation}
Substituting the aforementioned expression for $A_l$ into (\ref{eq:fourier_series}) and applying the usual Poisson summation formula yields
\begin{equation}
\label{eq:temptemp}
    \begin{split}
        \bar{A}(\theta)=&\sum_{m=-\infty}^{\infty}\int_{-\infty}^{\infty}\int_{0}^{z_0}
        (z_0-z)[K(z-xz_0)\\
        &+K(z+xz_0)]dze^{ix(\theta-2m\pi)}dx
    \end{split}
\end{equation}
Let $\bar{K}(\zeta)$ denote the Fourier transform (not to be confused with our similar notation for Fourier series) of $K(z)$. Eqn. (\ref{eq:temptemp}) can be rewritten as
\begin{equation}
\label{eq:atheta}
    \bar{A}(\theta)=z_0\sum_{m=-\infty}^{\infty}\bar{K}\left(\frac{2m\pi-\theta}{z_0}\right)\frac{\sin^2{\frac{\theta}{2}}}{\left(m\pi-\frac{\theta}{2}\right)^2}
\end{equation}
Integrals 2.5.25.9 and 2.5.25.15 of \cite{prudnikov} and the detailed analytic-continuation arguments of \cite{wu} give  $\overline{K}(\zeta)$ as
\begin{equation}
\label{eq:kernel_fourier}
 \overline{K}(\zeta)=
    \frac{1}{2\pi}K_0\left(a\sqrt{\zeta^2-k_c^2}\right)
\end{equation}
where $K_0$ is the modified Bessel function. Eqns. (\ref{eq:fourierexp}), (\ref{eq:btheta}), (\ref{eq:atheta}), and (\ref{eq:kernel_fourier}) provide the \textit{exact}
solution of the Toeplitz system (\ref{eq:toeplitz}) for \textit{nonzero} discretization length $z_0$. 

As $z_0\rightarrow 0$, (\ref{eq:kernel_fourier}) shows that $\overline{K}(\zeta)$ is exponentially small: 
\begin{equation}
\label{eq:kernel_asymptotic}
    \bar{K}(\zeta) \sim \frac{1}{2}\sqrt{\frac{1}{2\pi a|\zeta|}}e^{-a|\zeta|}
\end{equation}
We can thus approximate the denominator in the integrand of (\ref{eq:fourierexp}) by the first two terms ($m=0$ and $m=1$) in the sum (\ref{eq:atheta}) and then set $\phi=\pi-\theta$ to obtain (\ref{eq:big_1}), shown at the bottom of this page. As $z_0\rightarrow 0$, the dominant contribution to the integral in (\ref{eq:big_1}) comes from a small interval near $\phi=0$. Thus  the contribution $\int_1^{\pi}$ is negligible and we can change the upper integration limit in  (\ref{eq:big_1}) from $\pi$ to  1. We then replace the two instances of  $\bar{K}$ using (\ref{eq:kernel_asymptotic}) and set $x=\phi a/z_0$ to get

\newcounter{mytempeqncnt1}
\setcounter{mytempeqncnt1}{\value{equation}}% Set the equation number to one less than the one% desired for the first equation here.% The value here will have to changed if equations% are added or removed prior to the place these% equations are referenced in the main text.
\setcounter{equation}{18}

\begin{equation}
\label{eq:i_approximate_close}
    I^{(\infty)}_n \sim \frac{1}{\sqrt{2\pi}}k_cz_0\sqrt{\frac{z_0}{a}}(-1)^ne^{\pi\frac{a}{z_0}}
    f\left(\frac{z_0}{a},k_ca,n\frac{z_0}{a}\right)
\end{equation}
where $f$ is given by eqns. (35) and (36) of \cite{fikioris1} with $k$ and $\zeta$ replaced by $k_c$ and $\zeta_c$. Those two equations and steps identical to steps in \cite{fikioris1} allow us to obtain the first two terms in the Taylor-series expansion of $f$ in powers of $z_0/a$.  Upon substituting the expression for $f$ thus obtained into (\ref{eq:i_approximate_close}), we arrive at the desired asymptotic result (\ref{eq:asymptotic}).

\end{document}